\documentclass[aps,twocolumn,nofootinbib]{revtex4}
\usepackage{graphicx}
\usepackage{adjustbox}
\usepackage{dcolumn}
\usepackage{bm}
\usepackage{amsmath}
\usepackage{bbold}
\usepackage{multirow}
\usepackage{amsmath}
\usepackage[letterpaper, hmargin = {2cm}, vmargin = {2cm}]{geometry}
\usepackage{tablefootnote}

\usepackage[ulem=normalem]{changes}
\usepackage{mathrsfs}
\usepackage{float}

\begin{document}

\title{ The Core of  $^{25}$F in the Rotational Model}

\author{ A.~O.~Macchiavelli,  R.~M.~Clark,  H.~L.~Crawford,   P.~Fallon, I.~Y.~Lee,  C.~Morse, C.~M.~Campbell, M.~Cromaz, and C.~Santamaria}
\affiliation{Nuclear Science Division, Lawrence Berkeley National Laboratory, Berkeley, CA 94720, USA}
\date{\today}
 
\begin{abstract}
In a recent  experiment, carried out at RIBF/RIKEN,  the  $^{25}$F$(p,2p)$$^{24}$O reaction was studied at 270 MeV/A in inverse kinematics.  Derived spectroscopic factors suggest that the effective core of $^{25}$F significantly differs from a free $^{24}$O nucleus.
We interpret these results within the Particle-Rotor Model and show that the experimental level scheme of  $^{25}$F can be understood in the rotation-aligned coupling scheme, 
with  its $5/2^+_1$ ground state as the band-head of a decoupled band.   The excitation energies of the observed $1/2_1^+$ and  $9/2_1^+$ states correlate strongly with the rotational energy 
of the effective core, seen by the odd proton,  and allow us to estimate its  $2^+$ energy at  $\approx$ 3.2 MeV  and a moderate quadrupole deformation,  $\epsilon_2 \approx 0.15$.   The measured fragmentation of the $\pi d_{5/2}$ single-particle strength is discussed and some further experiments suggested.
  

\end{abstract}


\maketitle

\section{Introduction}

The structure of  neutron-rich nuclei is a central theme of study in the field of nuclear structure.  Of particular interest is the quest to  understand the evolution of shell-structure and collectivity with isospin.
The emergence of the Islands of Inversion at  $N$=8, 20, and 40 are prime examples of such evolution and have provided strong evidence 
regarding the important role played by the neutron-proton force~\cite{Otsuka20, Sor08, Poves, Ots01, Heyde1, Warburton90}.

Another intriguing and dramatic impact of the action of the neutron-proton force is seen in the so-called oxygen neutron-dripline anomaly, at $N=16$, which is extended to $N=22$ in for the F isotopes 
with just the addition of one $ d_{5/2}$ proton.  In a recent work~\cite{tang20},  the structure of $^{25}$F has been investigated  
via $(p,2p)$ quasi-free knockout experiments with exclusive
measurements using  a $^{25}$F  beam at   RIBF/RIKEN. %
The analysis of measured cross-sections 
and derived spectroscopic factors may imply that the core of $^{25}$F
consists of $\sim$ 35\% $^{24}$O$_{gs}$ and $\sim$ 65\%
excited  $^{24}$O.    As discussed by the authors, their results suggest that the addition of the $0d_{5/2}$ proton considerably changes 
the neutron structure in $^{25}$F from that in $^{24}$O, and calls for a revision to the $np$ tensor interaction in the
widely used effective interactions, which appears to be too weak to
reproduce the observations. In contrast,  studies of neutron decay from unbound excited states in $^{24}$O~\cite{Calem09} and one-neutron removal from  $^{24}$O~\cite{Ritu09} were indicative of a $N$=16 shell closure and the doubly-magic nature of this nucleus. The relatively high excitation energy $E_x = 4.7 \pm 0.1$ and the small  $B(E2) \approx 1/2$  WU (Weisskopf units) of the $2^+_1$ state~\cite{Tshoo12}  has further supported this interpretation.

In this article we follow up on our earlier work~\cite{aom} and interpret the above results in terms of
a collective picture,  within the framework of the Particle-Rotor Model (PRM)~\cite{Larsson, Rag} to provide further insight into the nature of the effective  $^{24}$O core  in  $^{25}$F.

\section{The Structure of $^{25}$F}

\begin{figure} 
\centering
\includegraphics[width=9cm,angle=0]{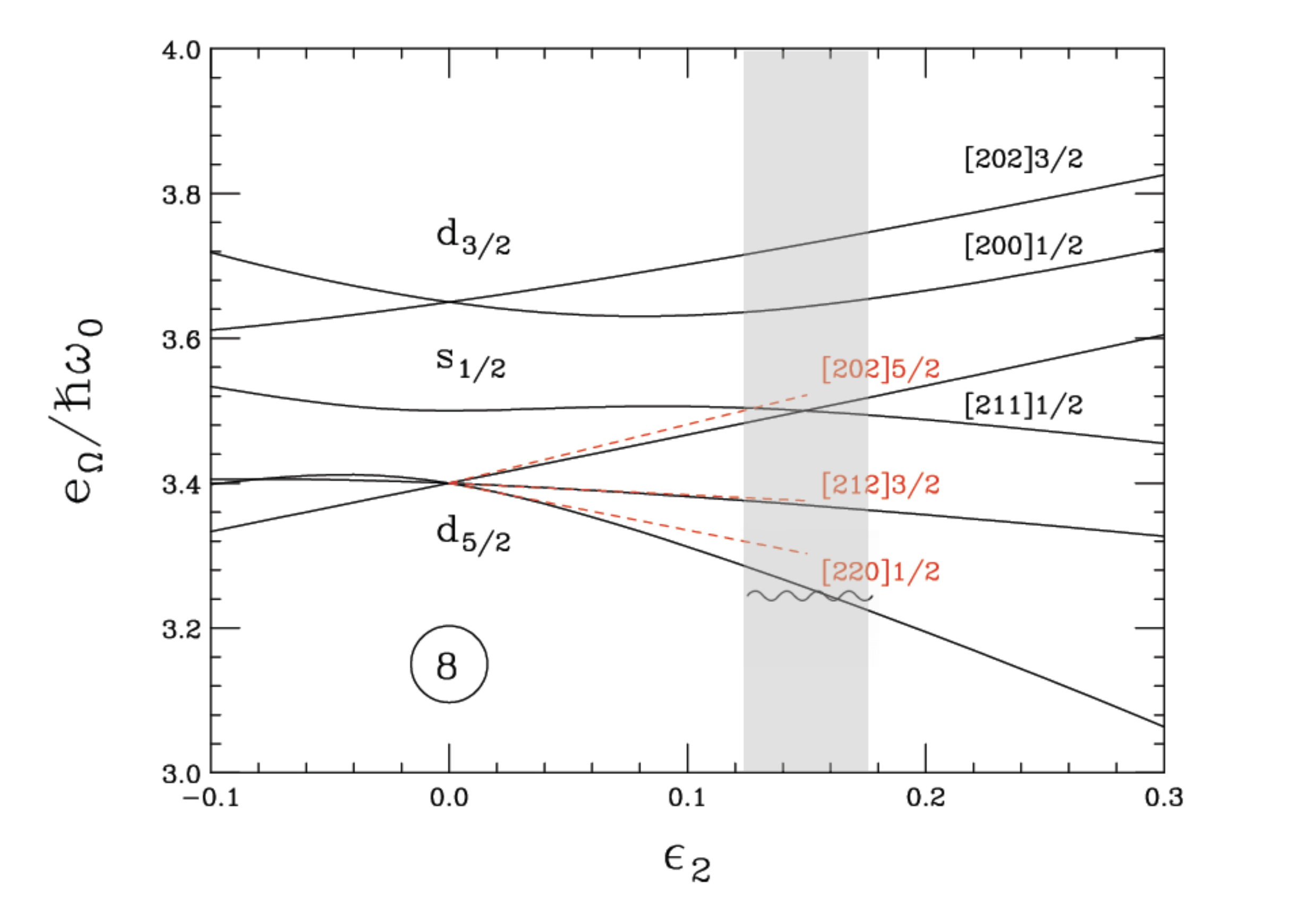}
\caption{Nilsson levels relevant for the structure of positive parity proton states in $^{25}$F, with the red dashed-lines 
representing  the single-$j$ approximation of the $d_{5/2}$ multiplet. The shaded area indicates the anticipated $\epsilon_2$ deformation and the wavy line  the Fermi level of the odd  proton.  Energies are in units of  the harmonic oscillator frequency, $\hbar\omega_0$. }
\label{fig-1}       
\end{figure}

The structure of odd-A nuclei usually
offers  fingerprints that can disentangle the competition of
single-particle and collective degrees of freedom if they
can be regarded, at least a priori, as one 
nucleon coupled to a core.   Considering $^{24}$O as our core, an inspection of the Nilsson diagram~\cite{Sven} in Fig.~\ref{fig-1} suggests that the odd proton will occupy the single-$j$ multiplet 
originating from the $d_{5/2}$ orbit, namely the levels  $[220]\frac{1}{2}$, $[211]\frac{3}{2}$, and $[202]\frac{5}{2}$, with its Fermi energy at the $\Omega=\frac{1}{2}$, as indicated by the wavy line in Fig.~\ref{fig-1}\\

The effects of rotation on the single-particle motion are well understood, and the Particle Rotor Model (PRM) has been very successful in explaining the observed near-{\sl yrast} structures  in deformed nuclei ~\cite{Frank2}.

The PRM Hamiltonian can be written as~\cite{Larsson,Rag}:
\begin{equation}
H=      H_p +   \frac{\hbar^{2}}{2\mathscr{I}}{\vec{R}^2}
\label{eq:eq1}
\end{equation}
where $H_p$ is the Nilsson Hamiltonian~\cite{Sven} for the particle in the absence
of rotation,  $ \mathscr{I}$ and $\vec{R} $ the moment of inertia and the angular momentum of the core respectively.
Replacing $\vec{R} = \vec{I} -\vec{j}$ in Eq.~\ref{eq:eq1}  gives the usual expression:
\begin{equation}
H=      E_\Omega +   \frac{\hbar^{2}}{2\mathscr{I}}I(I+1) + H_c
\label{eq:eq2}
\end{equation}
\noindent
where $E_\Omega$ are the intrinsic level energies and $H_c$ is the Coriolis coupling term

\begin{equation}
H_c=   - \frac{\hbar^{2}}{2\mathscr{I} }(I_+j_-+I_-j_+)
\label{eq:eq3}
\end{equation}
where $I_\pm$ and  $j_\pm$  are the ladder operators for the total and single particle angular momenta, respectively.  This coupling is particularly important  for small deformations and large $j$, and increases with the rotational frequency, $\omega_{rot}$. The Coriolis $K$-mixing gives rise to a wave-function of the general form
\begin{equation}
\psi_I  =  \sum_{K}  \mathcal{A}_{K}  | I K \rangle
\label{eq:eq4}
\end{equation}
 \begin{figure}
\centering
\includegraphics[width=8cm]{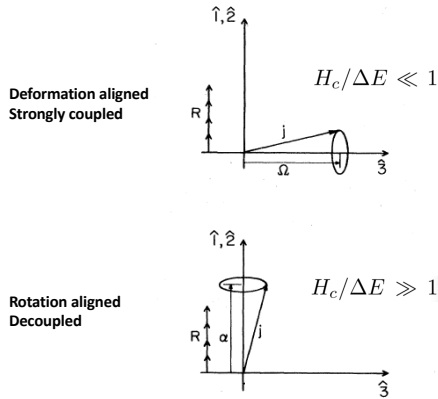}
\caption{Schematic representation of the strongly coupled and decoupled limits of the PRM.  The latter is used here  in our description of the  $^{25}$F. 
The symmetry axis is labeled $\hat{3}$. Collective rotation takes place around a perpendicular axis ($\hat{1}$, $\hat{2)}$.  Figure adapted from Ref.~\cite{Frank2}.}
\label{fig-2}       
\end{figure}
\noindent
The ratio of the Coriolis matrix elements  in Eq.~\ref{eq:eq3} ($H_c \sim\hbar^2Ij/\mathscr{I} \sim j \hbar\omega_{rot} $) to the intrinsic level spacings 
($\Delta E \sim \epsilon_2 \hbar\omega_0$) serves as a control parameter defining the characteristics of the coupling between collective and intrinsic angular momenta.
For $H_c/\Delta E \ll 1$, the particle remains strongly coupled to the core maintaining the projection of its angular momentum on the symmetry axis, $\Omega$,  as a good quantum number. 
When $H_c/\Delta E \gg 1$, a rotation-aligned coupling limit is anticipated~\cite{Frank1,Frank2}. In this case, the {\sl yrast} band  has spins $I=j,~j+2,~j+4,~...$,  and the energy spacings equal that of the core; this type of band is referred to as a decoupled band. The two limiting cases are illustrated in Fig.~\ref{fig-2}.  
\section{Results}

\subsection{Level energies}

\begin{figure}
\centering
\includegraphics[width=6cm]{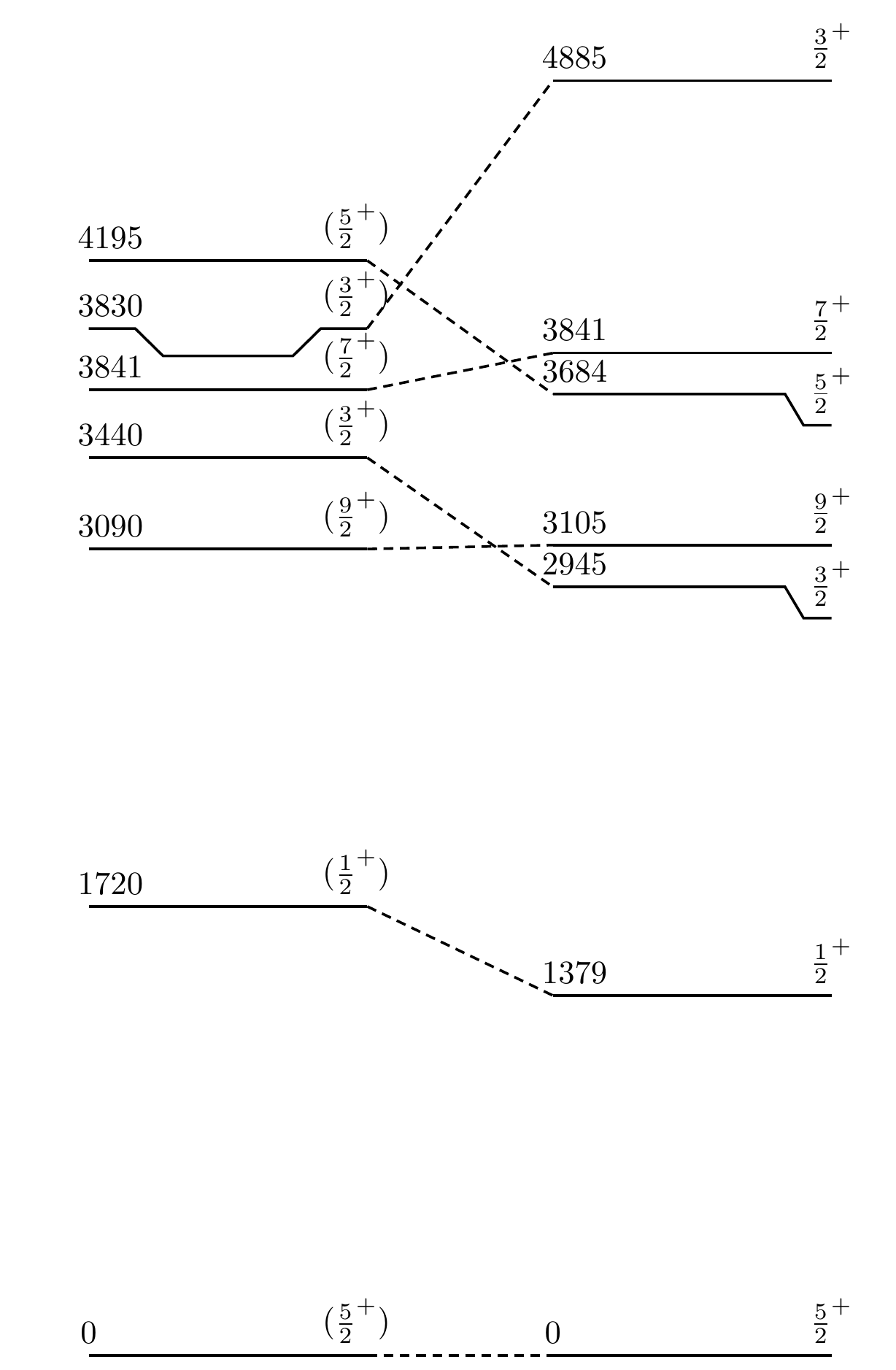}
\caption{ Left: the experimental level scheme of $^{25}$F from Ref.~\cite{Vajta14}. Right:  Results of the PRM calculations. Energies are in keV.}
\label{fig-3}        
\end{figure}
The experimental level scheme of $^{25}$F~\cite{Vajta14}, shown in Fig.~\ref{fig-3} (left side), exhibits an interesting pattern having the first two $yrast$ states with spins $5/2_1^+$ and $9/2_1^+$ and a conspicuous 
$1/2_1^+$  state in between.   In analogy with our interpretation in Ref.~\cite{aom} of the structure of $^{29}$F~\cite{Pieter}, the $yrast$ states can be associated with members of the decoupled band  based on the $d_{5/2}$ multiplet
and for which we have $(j \omega_{rot})/( \epsilon_2\omega_0) > 1$.  The $1/2_1^+$ must have anti-parallel coupling of $\vec{j}$ with the core rotation, $\vec{R}$.   It follows that in the decoupled limit ($\epsilon_2 \rightarrow 0$) the energy of the $1/2^+_1$ state with respect to the
ground state is proportional to the rotational energy of the core, $E_{2^+}(core)$.
Together with the $9/2_1^+$ state they provide a proxy for the $2^+$ energy of the effective $^{24}$O core in $^{25}$F.  Adjusting  to the energies of the $1/2_1^+$ and $9/2_1^+$ states gives  $E_{2^+}(core)\approx $ 3.2 MeV, in line with  a modest quadruple deformation, $\epsilon_2 \approx 0.15$, and consistent with the conditions required for the appearance of a decoupled band.

The results obtained of the PRM calculations, shown also in Fig.~\ref{fig-3} (right side) are in good agreement with the experimental data and give support to the rotational model description.   Furthermore, in the rotation-aligned coupling limit the amplitudes $\mathcal{A}_{K}$ entering in Eq.~\ref{eq:eq4} are given by the Wigner $\mathbb{d}$-function evaluated at $\pi/2$, the angle between the symmetry ($\hat{3}$)  and rotation axes ($\hat{1},\hat{2}$,)~\cite{Frank2}:
\begin{equation}
\mathcal{A}_{K} \approx \mathbb{d}^{5/2}_{5/2,K}(\pi/2)
\label{eq:eq5}
\end{equation}
\noindent
In  $^{26}$F~\cite{Lepa13},  the  $1^+$ ground and $4^+$ isomeric states 
 can be associated with the anti-parallel  and parallel couplings of the odd-neutron, in the $d_{3/2}$ Nilsson multiplet, to the structure of  $^{25}$F.   The former, favored by the Gallagher-Moszkowski rule~\cite{GM58},  gives a $1^+$ as the lowest state and the latter a  $4^+$ as the bandhead of a doubly-decoupled band. 
\subsection{Spectroscopic factors}

We now proceed with the calculation of spectroscopic factors for the $(-1p)$ knockout  reaction and compare them to those reported in Ref.~\cite{tang20}.  
Following the formalism discussed in Ref.~\cite{Elbek}, which we recently applied to a similar case in $^{18,19}$F~\cite{aom2},   we obtain the expression:
\begin{equation}
S_{i, f} (j\ell) = \big( \sum_{K}  \mathcal{A}_{K} \theta_{i, f}(j\ell,K) \big)^2
\label{eq:eq6}
\end{equation}
\begin{equation} 
\begin{split}
\theta_{i, f  }(j\ell,K)& =  \sqrt2 \langle I_{i}j  K \Omega_\pi  | I_{f} 0\rangle C_{j,\ell} \langle\phi_f|\phi_i\rangle 
\end{split}
\label{eq:eq7}
\end{equation}
where  $ \mathcal{A}_{K}$ are given in Eq.~\ref{eq:eq5},    $\langle | \rangle$ is a Clebsch-Gordan coefficient  and $\langle\phi_f|\phi_i\rangle$ represents the core overlap between the initial and final states, typically assumed to be 1.  Since we are considering a single-$j$ approximation for 
$d_{5/2}$ Nilsson multiplet, the amplitudes $C_{j,\ell}$ are equal to 1. Special care should be taken to assure consistency between the relative phases of the $\mathcal{A}_{K}$ Coriolis-mixed amplitudes  and the Clebsch-Gordan coefficients entering in the sum.  
 
In Table~\ref{table1}, the spectroscopic factors  are compared to the measurement reported in Ref.~\cite{tang20}.  The PRM is able to explain the level scheme of  $^{25}$F but predicts a small fragmentation of the $d_{5/2}$ proton strength, with $\approx$ 15\% going to the $2^+$ and $4^+$ of $^{24}$O (PRM1).    However,  the premise of a substantial difference between the initial and final cores requires that the overlap in Eq.~\ref{eq:eq7} should be considered explicitly.  
\begin{table}[ht]
\centering
\caption{  Comparison of the measured spectroscopic factors to the PRM results, with (PRM1) and without core overlap (PRM2).  Also shown are shell-model calculations using the SDPF-MU interaction.}
\bigskip
\begin{tabular}{c|c|cccc}
\hline\hline
Final State &  $S_{exp}$  &  &&  $S_{th}$ &  \\
in $^{24}$O &  Ref.~\cite{tang20} &    &PRM1& PRM2 & SDPF-MU \\
\hline 
Ground   &~0.36(13)~ & & 0.85 &  0.56&  0.95 \\
Excited &~0.65(25)~ & & 0.15 &  0.44&  0.05\\
\hline\hline
\end{tabular}
\label{table1}
\end{table}
We use the method described in Refs.~\cite{TNA1, TNA2} and obtain\footnote[1]{ A simple volume overlap gives $ \langle\phi_f|\phi_i\rangle \approx  \frac{1}{1+ \epsilon_2+ \frac{2}{3}\epsilon_2^2 +~ ...} \sim0.85$}  $\langle\phi_f|\phi_i\rangle \approx $ 0.81 bringing the PRM  result closer to the observations (PRM2).

For reference we also include the  shell-model results using the SDPF-MU interaction given in~\cite{tang20}.  Note, if the authors of Ref.~\cite{tang20} had corrected the gs to gs spectroscopic factor by a quenching factor $\sim$ 0.6, usually observed in $(p,2p)$ reactions~\cite{leila}, the agreement would have been excellent.  

Obviously, additional studies of proton addition and removal reactions will be of interest,  specifically proton knockout or $(d,^3He)$ from $^{26}$Ne come to mind. Here, anticipating that $\langle\phi_f|\phi_i\rangle \sim 1$, we predict an spectroscopic factor for the $0^+$  to $5/2^+$  transition  $S_{if} \approx 1.25$.

\section{Conclusion}
The rotational model is able to describe the structure of $^{25}$F as arising from the coupling of a proton $d_{5/2}$ Nilsson multiplet 
to an effective core of modest deformation,  $\epsilon_2 \sim 0.15$.  These conditions anticipate that the development of a decoupled band 
should be favorable and indeed,  PRM calculations show that the rotation aligned coupling scheme is in agreement  
with the observed levels.  Using the formalism developed for studies of single-nucleon transfer reactions in deformed nuclei,
we calculated the proton spectroscopic factors for the $^{25}$F($5/2^+) (-1p) ^{24}$O reaction.  Agreement with the experimental 
data~\cite{tang20} is obtained by the fragmentation of the $d_{5/2}$ strength due to both deformation and a core overlap.


The Nilsson plus PRM picture suggests that the extra proton, with a dominant component in the down-sloping $\frac{1}{2}[220]$ level,   polarizes $^{24}$O and stabilizes its dynamic deformation.
Thus, the  effective 
core in $^{25}$F can be interpreted as  a slightly deformed rotor with $E_{2^+}(core)\approx$ 3.2 MeV and  $\epsilon_2 \approx0.15$,
 compared to  the real doubly-magic $^{24}$O with $E_{2^+}\approx$ 4.7 MeV and weak vibrational quadrupole collectivity.  

Furthermore,  electromagnetic observables for the three lowest experimental levels obtained in the PRM (Table~\ref{Table 2}),
suggest that measurements of the magnetic and quadrupole moments of the $5/2^+$ state as well as a Coulomb excitation measurement of the transition probabilities, will definitely shed further light on the validity of our interpretation. 
 \begin{table}[ht]
\centering
\caption{ Electromagnetic properties of the low-lying levels of $^{25}$F in the PRM. Magnetic moments have been calculated using  $g_R=Z/A$ and  $g_s=0.7 (g_s)_{free}$.}
\bigskip
\begin{tabular}{c|c|c|c|c}
\hline\hline
$I^\pi$ &  $E_{x}$ & $\mu$        & $Q$               &   $B(E2;I \rightarrow \frac{5}{2}^+$)\\
           &   ~ [MeV]  ~ &~ [$\mu_N$] ~  & $ ~[efm^2]$ ~  & [WU]\\
 \hline 
$\frac{5}{2}^+$&   0  &  3.9 & -4.5 & ---\\
$\frac{1}{2}^+$& 1.4 &  1.9 &   0   &3.9\\     
$\frac{9}{2}^+$& 3.1 &  4.6 & -7.8 &1.9\\
                  \hline\hline
\end{tabular}
\label{Table 2}
\end{table}

 \begin{acknowledgments} 
This material is based upon work supported by the U.S. Department of Energy, Office of Science, Office of Nuclear Physics under Contract No. DE-AC02-05CH11231.  We would like to thank K.~Wimmer and R.~Kanungo for their comments on the manuscript.
\bigskip
\bigskip

\end{acknowledgments}

\end{document}